\documentclass[prl,twocolumn,showpacs]{revtex4}
\usepackage{graphicx,amsfonts,amssymb,amsmath}

\begin{document}

\title{Relation Between Bulk and Interface Descriptions of Alloy Solidification}

\date{\today}

\author{Alexander L. Korzhenevskii}
\affiliation{Institute for Problems of Mechanical Engineering,
RAS, Bol'shoi prosp. V. O., 61, St Petersburg, 199178, Russia}
\author{Richard Bausch}
\affiliation{Institut f{\"u}r Theoretische Physik IV,
 Heinrich-Heine-Universit{\"a}t D{\"u}sseldorf, Universit{\"a}tsstrasse
 1, D-40225 D{\"u}sseldorf, Germany}
\author{Rudi Schmitz}
\affiliation{Institut f{\"u}r Theoretische Physik C, RWTH Aachen University,
Templergraben 55, D-52056 Aachen, Germany}

\date{\today}

\begin{abstract}

From a simple bulk model for the one-dimensional steady-state solidification of a dilute binary alloy we derive an interface description, allowing arbitrary values of the growth velocity. Our derivation leads to exact expressions for the fluxes and forces at the interface and for the set of Onsager coefficients. We, moreover, discover a continuous symmetry, which appears in the low-velocity regime, and there deletes the Onsager sign and symmetry properties. An example is the generation of the sometimes negative friction coefficient in the crystallization flux-force relation.

\end{abstract}


\maketitle

\section{Introduction}

The interface kinetics at the solidification front of a growing dilute binary alloy has been described by Baker, and Cahn \cite{Baker-Cahn} within the framework of linear irreversible thermodynamics. An apparent violation of the Onsager symmetries in their approach has led to several partly controversial discussions in papers by Caroli, Caroli, and Roulet \cite{Caroli}, by Kaplan, Aziz, and Gray \cite{Kaplan}, and in the review \cite{Hillert} by Hillert. Another apparent paradox is the possible appearance of a negative friction coefficient in the crystallization flux-force relation. This unexpected effect has been observed in the phase-field approach by Karma and Rappel \cite{Karma}, and in Ref. \cite{PRE-1} has been shown to be compatible with the principle of a positive entropy production. A thorough discussion of this point is also contained in a recent phenomenological approach to the one-dimensional steady-state solidification by Brener and Temkin \cite{Brener-Temkin}.

In the present work we derive the interface description for the solidification of a dilute binary alloy from a simple bulk model. The latter is a conveniently chosen specific version of the capillary-wave model, derived in Ref. \cite{PRE-1} from a phase-field model. As in Ref. \cite{Brener-Temkin}, we consider the scenario of a one-dimensional steady-state motion of the solidification front, however, without any restriction on the value of the growth velocity. Our main results include exact expressions for the fluxes and driving forces at the interface, carrying the full velocity dependence due to the solute trapping effect. We also obtain exact expressions for the Onsager coefficients, which, in the basis of crystallization and diffusion fluxes and forces, turn out to be independent of the growth velocity.

Another result of our investigations is the observation that a continuous symmetry arises in the low-velocity regime due to a linear dependence between the two fluxes of the model. This symmetry conserves the transport equations through the interface and, as a consequence, the entropy production within the interface region, but generates two independent redundant parameters in the set of transport coefficients. In general, this effect deletes the Onsager sign and symmetry conditions of this set, and, by properly choosing the parameters, leads to the appearance of a sometimes negative friction coefficient in the crystallization flux-force relation. As to be expected, this apparent anomaly can easily be seen not to violate the principle of a positive entropy production.

\section{Capillary-Wave Model}

Based on a phase-field description, we have established in Ref. \cite{PRE-1} a capillary-wave model for the solidification of a dilute binary alloy where the interface position $Z({\bf x},t)$ enters as an extra field variable in addition to the solute concentration $C({\bf r},t)$. In the one-dimensional case the effective Hamiltonian of our model reads

\begin{equation}\label{Hamiltonian}
H=\frac{\kappa}{2}\int_{-\infty}^{+\infty}dz[C(z,t)-U(z-Z(t))]^2\,\,.
\end{equation}
Here, in addition to the coupling constant $\kappa$, the function $U(z-Z)$ enters as an input quantity, which, following from the condition $\delta H/\delta C=0$, means the equilibrium concentration of the solute. The equations of motion of the model are given by

\begin{eqnarray}\label{dynamics}
&&\partial_t Z=-\Lambda\frac{\delta H}{\delta Z}\,\,,\\
&&\partial_t C=\partial_z D(z-Z(t))\partial_z\frac{1}{\kappa}\frac{\delta H}{\delta C}\,\,,\nonumber
\end{eqnarray}
where $\Lambda$ and $D(z-Z)$ are kinetic coefficients, measuring the mobility of the interface and the local diffusivity of the solute. Instead of including an external driving force in the first of these equations, as done in Ref. \cite{PRE-1}, the solidification process will be activated, as in Ref. \cite{Brener-Temkin}, by imposing appropriate boundary conditions on the second of the equations (\ref{dynamics}).

In a steady-state solidification process with constant growth velocity $V$ in $z$-direction, the solute concentration in the comoving frame only depends on the variable

\begin{equation}\label{comoving-frame}
\zeta\equiv z-Vt\,\,.
\end{equation}
As a result, insertion of Eq. (\ref{Hamiltonian}) into the first equation in Eqs. (\ref{dynamics}) yields

\begin{equation}\label{force-balance}
V=-\Lambda\kappa\int_{-\infty}^{+\infty}d\zeta U'(\zeta)\left[C(\zeta)-U(\zeta)\right]\,\,.
\end{equation}
The second equation in Eqs. (\ref{dynamics}) can be integrated once, which, with the boundary condition

\begin{equation}\label{boundary}
C(-\infty)=C_0\,\,,
\end{equation}
leads to the first-order differential equation

\begin{equation}\label{integr-diff}
[C_0-C(\zeta)]V=D(\zeta)[C'(\zeta)-U'(\zeta)]\,\,.
\end{equation}

Eq. (\ref{force-balance}) can be rewritten in form of the force-balance equation

\begin{equation}\label{balance}
-\frac{V}{\Lambda}+F+G(V)+\kappa\frac{(\Delta C)^2}{2}=0
\end{equation}
where the first term has the form of a viscous-friction force. The contribution

\begin{equation}\label{driving-force}
F\equiv\kappa(\Delta C)(C_S-C_0)
\end{equation}
acts as an external driving force where, in terms of the solute concentrations

\begin{equation}\label{liq-sol-conc}
C_S\equiv U(-\infty)\,\,\,,\,\,\,C_L\equiv U(+\infty)
\end{equation}
in the liquid and solid phases,

\begin{equation}\label{miscibility gap}
\Delta C\equiv C_L-C_S
\end{equation}
means the miscibility gap. Evidently, $F$ describes the effect of a quench from the equilibrium value $C_S$ to some non-equilibrium value $C_0$, which for $C_S-C_0\ge 0$ induces a growth rate $V\ge 0$. The contribution

\begin{equation}\label{drag-force}
G(V)\equiv-\kappa\int_{-\infty}^{+\infty}d\zeta U'(\zeta)[C(\zeta)-C_0]
\end{equation}
is a drag force, generated by the comoving solute layer at the interface. Using the steady-state boundary condition

\begin{equation}\label{steady}
C(+\infty)=C(-\infty)\,\,,
\end{equation}
it can be rewritten, due to Eq. (\ref{integr-diff}), in the form

\begin{equation}\label{drag}
G(V)\equiv-\kappa\int_{-\infty}^{+\infty}d\zeta \frac{D(\zeta)}{V}[U'(\zeta)-C'(\zeta)]^2\,\,,
\end{equation}
which, for velocities $0\le V<\infty$, is negative and shows the appropriate behavior $G(V)\rightarrow 0$ for $V\rightarrow\infty$. At $V=0$, corresponding to the choice $C_0=C_S$, Eq. (\ref{balance}) implies $G(0)=-\kappa(\Delta C)^2/2$, elucidating the role of the last term in Eq. (\ref{balance}).

Ref. \cite{PRE-1} also provides an expression for the density $p(\zeta)$ of the entropy production. The phase field $\Phi(\zeta)$, entering this expression, is, according to Ref. \cite{PRE-1}, related to $U(\zeta)$ by the equation

\begin{equation}\label{phase-field}
\Phi(\zeta)\equiv\frac{2}{\Delta C}U(\zeta)-\frac{C_L+C_S}{\Delta C}
\end{equation}
and obeys the normalization condition

\begin{equation}\label{normalization}
\int_{-\infty}^{+\infty}\frac{d\zeta}{\xi}\left[\xi\partial_\zeta\Phi(\zeta)\right]^2=1
\end{equation}
where $\xi$ is a microscopic length scale of the order of the interface thickness. Due to Eq. (\ref{phase-field}), the related passage in the discussion of Ref. \cite{PRE-1} leads to the obviously positive expression

\begin{eqnarray}\label{S-production}
p(\zeta)&=&\frac{1}{T}\frac{V^2}{\Lambda}\left(\frac{2}{\Delta C}\right)^2\frac{1}{\xi}\left[\xi\partial_\zeta U(\zeta)\right]^2
\nonumber\\&+&\frac{1}{T}\frac{D(\zeta)}{\kappa}\left(\partial_\zeta\frac{\delta H}{\delta C(\zeta)}\right)^2\,\,.
\end{eqnarray}
for the density of the entropy production at some fixed temperature $T$.

In order to explicitly evaluate Eqs. (\ref{integr-diff}), (\ref{balance}), and (\ref{S-production}), we now consider the specific model, defined by

\begin{eqnarray}\label{equil-conc}
&&U(\zeta)=C_S+\frac{1}{4\xi}(C_L-C_S)\\&&\nonumber\\&&\times
\left\{
\begin{array}{c}
0\,\,\,\,\,\,\,\,\,\,\,\,\,\,\,\,\,\,\,\,\,\,\,\,\,\,\,\,\,\,\,\,\,\,\,\,\,\,\,\,\,\,\,\,\,\,\,\,\,\,
\,\,\,\,\zeta<-2\xi\\{\ }\\\zeta+2\xi\,\,\,\,\,\,\,\,\,\,\,\,\,\,\,\,\,\,\,\,\,\,\,\,\, \,\,\,\,\,\,\,\,\,\,-2\xi<\zeta<+2\xi\\{\ }\\4\xi\,\,\,\,\,\,\,\,\,\,\,\,\,\,\,\,\,\, \,\,\,\,\,\,\,\,\,\,\,\,\,\,\,\,\,\,\,\,\,\,\,\,\,\,\,\,\,\,\,+2\xi<\zeta\,\,,
\end{array}
\right.
\nonumber
\end{eqnarray}

\begin{eqnarray}\label{diff-coeff}
&&D(\zeta)=D_S+\frac{1}{4\xi}(D_L-D_S)\\&&\nonumber\\&&\times
\left\{
\begin{array}{c}
0\,\,\,\,\,\,\,\,\,\,\,\,\,\,\,\,\,\,\,\,\,\,\,\,\,\,\,\,\,\,\,\,\,\,\,\,\,\,\,\,\,\,\,\,\,\,\,\,\,\,
\,\,\,\,\zeta<-2\xi\\{\ }\\\zeta+2\xi\,\,\,\,\,\,\,\,\,\,\,\,\,\,\,\,\,\,\,\,\,\,\,\,\, \,\,\,\,\,\,\,\,\,\,-2\xi<\zeta<+2\xi\\{\ }\\4\xi\,\,\,\,\,\,\,\,\,\,\,\,\,\,\,\,\,\, \,\,\,\,\,\,\,\,\,\,\,\,\,\,\,\,\,\,\,\,\,\,\,\,\,\,\,\,\,\,\,+2\xi<\zeta\,\,.
\end{array}
\right.
\nonumber
\end{eqnarray}
Here, $D_S$ and $D_L$ are the diffusion coefficients of the solute in the solid and liquid phases, and the width  $4\xi$ of the well-encompassed interface region has been chosen to ensure the normalization (\ref{normalization}). In the special case $D_S=0$ the solution of Eq. (\ref{integr-diff}), subject to the boundary condition (\ref{steady}), has the simple form

\begin{eqnarray}\label{C(zeta)}
&&C(\zeta)=C_0+\frac{1}{4\xi}(C_L-C_S)\left(1+4\xi\frac{V}{D_L}\right)^{-1}\\&&\nonumber\\&&\times
\left\{
\begin{array}{c}
0\,\,\,\,\,\,\,\,\,\,\,\,\,\,\,\,\,\,\,\,\,\,\,\,\,\,\,\,\,\,\,\,\,\,\,\,\,\,\,\,\,\,\,\,\,\,\,\,\,\,\,
\,\,\,\,\,\,\,\zeta<-2\xi\\{\ }\\(\zeta+2\xi)\,\,\,\,\,\,\,\,\,\,\,\,\,\,\,\,\,\,\,\,\,\,\,\,\, \,\,\,\,\, \,\,\,\,\,-2\xi<\zeta<+2\xi\\{\ }\\4\xi\exp{\left[-\frac{V}{D_L}(\zeta-2\xi)\right]}\,\,\,\,\,\,\,+2\xi<\zeta\,\,,
\end{array}
\right.
\nonumber
\end{eqnarray}
showing, due to the factor $(1+4\xi V/D_L)^{-1}$, the solute-trapping effect. Some properties of this solution have already been discussed in Ref. \cite{PRE-1}, and below will be used to establish the interface description of the solidification process.

The entropy production $P$, generated in the interface region per unit area, follows from Eq. (\ref{S-production}) by integration over the interface region $-2\xi<\zeta<+2\xi$. Taking into account Eqs. (\ref{Hamiltonian}), (\ref{integr-diff}), (\ref{balance}), (\ref{drag-force}), and (\ref{normalization}), one obtains

\begin{eqnarray}\label{S-prod}
P&\equiv&\frac{V}{T}\left[F+G(V)+\frac{\kappa}{2}(\Delta C)^2\right]\\&+&
\frac{V}{T}\int_{-2\xi}^{+2\xi}d\zeta[C_0-C(\zeta)]\kappa[C'(\zeta)-U'(\zeta)]\nonumber\\&=&
\frac{V}{T}\left[F+\frac{\kappa}{2}(\Delta C)^2-\frac{\kappa}{2}(C_+-C_-)^2\right]\nonumber
\end{eqnarray}
where we have used the notation

\begin{equation}\label{plus-minus}
C_{\pm}\equiv C(\zeta=\pm 2\xi)\,\,,
\end{equation}
and the identity

\begin{equation}\label{C_-=C_0}
C_-=C_0\,\,,
\end{equation}
implied by Eq. (\ref{C(zeta)}). By insertion of the solution (\ref{C(zeta)}) into the force balance (\ref{balance}) one, furthermore, finds the relation

\begin{equation}\label{C_0(V)}
C_0=C_S-(\Delta C)\left(\frac{1}{V_C}-\frac{1}{V_D+2V}\right)V
\end{equation}
written in terms of crystallization and diffusion velocities

\begin{equation}\label{velocities}
V_C\equiv\kappa(\Delta C)^2\Lambda\,\,\,,\,\,\,V_D\equiv\frac{D_L}{2\xi}\,\,.
\end{equation}
According to the result (\ref{C_0(V)}), the limit $F\rightarrow 0$ of the driving force (\ref{driving-force}) not only is compatible with the behavior $V\rightarrow 0$, but also with the possibility $V\rightarrow (V_C-V_D)/2$. In the process of establishing the interface description, this observation leads us to take care of the full velocity dependence, appearing in the solution (\ref{C(zeta)}).

\section{Interface Description}

The last expression in Eq. (\ref{S-prod}) can be written in the standard form

\begin{equation}\label{S-CD}
P=I_CE_C+I_DE_D
\end{equation}
where the crystallization and diffusion fluxes $I_C,I_D$ and the related forces $E_C,E_D$ are exactly given by

\begin{eqnarray}\label{CD-force-flux}
&&I_C\equiv\rho V\,\,,\\
&&E_C\equiv\frac{1}{T}\frac{F}{\rho}\,\,,\nonumber\\
&&I_D\equiv\frac{1}{2}(C_--C_+-\Delta C)V\,\,,\nonumber\\
&&E_D\equiv\frac{\kappa}{T}(C_+-C_--\Delta C)\,\,,\nonumber
\end{eqnarray}
$\rho$ meaning the density of the material. An equivalent but more familiar expression for $E_D$ is

\begin{equation}\label{E_D}
E_D=\frac{\delta \mu}{T}\equiv\frac{1}{T}[\mu(+2\xi)-\mu(-2\xi)]
\end{equation}
where $\mu(\zeta)$ is the chemical potential, defined by

\begin{equation}\label{chem-pot}
\mu(\zeta)\equiv\frac{\delta H}{\delta C(\zeta)}=\kappa[C(\zeta)-U(\zeta)]\,\,.
\end{equation}

The constitutive equations, connecting the fluxes and forces (\ref{CD-force-flux}), are obtained by evaluating Eqs. (\ref{integr-diff}) and (\ref{balance}) by means of Eqs. (\ref{equil-conc}), (\ref{diff-coeff}), and (\ref{C(zeta)}). This leads to the result

\begin{equation}\label{constit-CD}
\left( \begin{array}{ccc}
E_C \\\ E_D \\
\end{array} \right)=
\frac{\kappa}{TV_D}\left( \begin{array}{ccc}
R_{CC} &R_{CD} \\\ R_{DC} &R_{DD} \\
\end{array} \right)
\left( \begin{array}{ccc}
I_C \\\ I_D \\
\end{array} \right)
\end{equation}
with the again exact expressions of the matrix elements

\begin{eqnarray}\label{elements-CD}
&&R_{CC}=\left(\frac{V_D}{V_C}+1\right)\left(\frac{\Delta C}{\rho}\right)^2\,\,\,,\,\,\,R_{DD}=4\,\,,\nonumber\\
&&R_{CD}=2\frac{\Delta C}{\rho}\,\,\,\,,\,\,\, R_{DC}=2\frac{\Delta C}{\rho}\,\,.
\end{eqnarray}

We notice that the fluxes and forces (\ref{CD-force-flux}) carry the full velocity dependence, reflecting the solute trapping effect whereas the kinetic coefficients (\ref{elements-CD}) turn out to be independent of the velocity. The latter, moreover, obey the basic properties

\begin{eqnarray}\label{Onsager}
&&R_{CD}=R_{DC}\,\,\,,\,\,\,R_{CC}\ge 0\,\,\,,\,\,\,R_{DD}\ge 0\,\,,\nonumber\\
&&R_{CC}R_{DD}-R_{CD}^2\ge 0
\end{eqnarray}
of general Onsager coefficients.

In principle, the results (\ref{CD-force-flux}), (\ref{constit-CD}), (\ref{elements-CD}) are of limited interest, since the solution (\ref{C(zeta)}), in combination with the resulting explicit form of the force balance (\ref{balance}), describe the complete steady-state behavior of our model in the interface and bulk regions. However, these results are a good starting point for an exhaustive discussion of the interface description in the often considered low-velocity regime.

\section{Low-Velocity Limit}

Whereas the fluxes, defined in Eqs. (\ref{CD-force-flux}), are linearly independent for arbitrary values of $V$, one finds in linear order in $V$ the connection

\begin{equation}\label{low-V}
I_D=-\frac{\Delta C}{\rho}I_C\,\,.
\end{equation}
This relation implies the existence of a two-parameter continuous symmetry, which changes the coefficients (\ref{elements-CD}), but conserves the kinetic equations (\ref{constit-CD}), and, as a consequence, also the entropy production (\ref{S-CD}). In fact, the transformation

\begin{equation}\label{transformation}
\widetilde{\bf R}={\bf R}+{\bf S}\,\,,
\end{equation}
involving the matrices

\begin{equation}\label{R-matrix}
{\bf R}=\left( \begin{array}{ccc}
R_{CC} &R_{CD} \\\ R_{DC} &R_{DD} \\\end{array} \right)\,\,\,,\,\,\,{\bf S}=\left( \begin{array}{ccc}
S_{CC} &S_{CD} \\\ S_{DC} &S_{DD} \\\end{array} \right)\,\,,
\end{equation}
with coefficients

\begin{eqnarray}\label{S-elements}
&&S_{CC}=\lambda\left(\frac{\Delta C}{\rho}\right)^2\,\,\,,\,\,\,S_{DD}=\mu\,\,,\nonumber\\&&S_{CD}=\lambda\frac{\Delta C}{\rho}\,\,\,,\,\,\,
S_{DC}= \mu\frac{\Delta C}{\rho}\,\,,
\end{eqnarray}
leads, due to Eq. (\ref{low-V}), to separate cancellations of the contributions containing the free parameters $\lambda$ and $\mu$ in Eq. (\ref{constit-CD}). However, the matrix elements (\ref{elements-CD}) transform into the $\lambda,\mu$-dependent coefficients

\begin{eqnarray}\label{R-transf}
&&\widetilde R_{CC}=\left(\frac{V_D}{V_C}+\lambda+1\right)\left(\frac{\Delta C}{\rho}\right)^2\,\,\,,\,\,\,\widetilde R_{DD}=\mu+4\,\,,\nonumber\\
&&\widetilde R_{CD}=(\lambda+2)\frac{\Delta C}{\rho}\,\,\,,\,\,\,\widetilde R_{DC}=(\mu+2)\frac{\Delta C}{\rho}\,\,.
\end{eqnarray}
We mention that this result is in accordance with an old observation by Hooyman and De Groot \cite{De Groot}. It should also be noted that in the subdomain

\begin{equation}\label{subdomain}
\lambda=\mu\ge 0
\end{equation}
the coefficients (\ref{R-transf}) obey the Onsager conditions (\ref{Onsager}) in a formal sense. In this domain they depend, however, on the redundant parameter $\lambda$.

Generally, it is fairly allowed to ignore the constraint (\ref{subdomain}) and to choose in Eqs. (\ref{R-transf}) the values

\begin{equation}\label{diagonal}
\lambda=\mu=-2\,\,,
\end{equation}
which leads to a diagonal form of the matrix $\widetilde{\bf R}$ and, consequently, to the decoupled transport equations

\begin{eqnarray}\label{neg-fric}
&&E_C=\frac{\kappa}{T}\left(\frac{\Delta C}{\rho}\right)^2\left(\frac{1}{V_C}-\frac{1}{V_D}\right)I_C\,\,,\nonumber\\ &&E_D=\frac{\kappa}{T}\frac{2}{V_D}I_D\,\,,
\end{eqnarray}
explicitly showing that the coefficient, connecting $E_C$ and $I_C$ can become negative. Due to Eqs. (\ref{S-CD}) and (\ref{low-V}) the related entropy production is, in the presently discussed low-velocity regime, given by the expression

\begin{equation}\label{P}
P=\frac{\kappa}{T}\left(\frac{\Delta C}{\rho}\right)^2\left(\frac{1}{V_C}+\frac{1}{V_D}\right)I_C^2\,\,,
\end{equation}
which obviously is positive.

\section{Two-Component Representation}

The fluxes, associated with the two components of the solidifying alloy, are given by the relations

\begin{equation}\label{AB-fluxes}
\left( \begin{array}{ccc}
J_A \\\ J_B \\
\end{array} \right)=
\left( \begin{array}{ccc}
1-X &-1 \\\ X &1 \\
\end{array} \right)
\left( \begin{array}{ccc}
I_C \\\ I_D \\
\end{array} \right)\,\,,
\end{equation}

\begin{equation}\label{AB-forces}
\left( \begin{array}{ccc}
F_A \\\ F_B \\
\end{array} \right)=
\left( \begin{array}{ccc}
1 &\,&-X \\ 1 &\,&1-X \\
\end{array} \right)
\left( \begin{array}{ccc}
E_C \\\ E_D \\
\end{array} \right)\,\,,
\end{equation}
where, by definition,

\begin{equation}\label{C-X}
X\equiv\frac{1}{2\rho}(C_-+C_++\Delta C)\,\,.
\end{equation}
In the entropy production

\begin{equation}\label{S-AB}
P=F_AJ_A+F_BJ_B\,\,,
\end{equation}
arising from the similar expression (\ref{S-CD}), all $X$-dependent terms cancel. The special choice (\ref{C-X}), however, leads to the relations

\begin{eqnarray}\label{2-components}
&&J_A=(\rho-C_-)V\,\,\,,\,\,\,J_B=C_-V\,\,,\\&&F_A=\frac{1}{T}[\mu_A(+2\xi)-\mu_A(-2\xi)]\,\,\nonumber
\end{eqnarray}
where, parallel to Eq. (\ref{chem-pot}),

\begin{equation}\label{mu_A}
\mu_A(\zeta)\equiv\frac{1}{\rho}\left[1-C(\zeta)\frac{\partial}{\partial C(\zeta)}\right]\frac{\kappa}{2}[C(\zeta)-U(\zeta)]^2\,\,,
\end{equation}
which, together with Eq. (\ref{chem-pot}), provides agreement with standard notations. It is obvious that, in discussions of the low-velocity limit, one would replace the quantity (\ref{C-X}) by the approximate value $X=(\Delta C)/\rho$.

which together with Eq. (\ref{chem-pot}) establishes agreement with the notions, used in Ref. \cite{Baker-Cahn}. It is obvious that, in a discussion of the low-velocity limit, one would replace the quantity (\ref{C-X}) by the approximate value $X=(\Delta C)/\rho$.

The constitutive equations (\ref{constit-CD}), finally, transform into

\begin{equation}\label{constit-AB}
\left( \begin{array}{ccc}
F_A \\\ F_B \\
\end{array} \right)=
\frac{\kappa}{TV_D}\left( \begin{array}{ccc}
Q_{AA} &Q_{AB} \\\ Q_{BC} &Q_{BB} \\
\end{array} \right)
\left( \begin{array}{ccc}
J_A \\\ J_B \\
\end{array} \right)
\end{equation}
where the kinetic coefficients in this basis are related to those in the $C,D$ basis by

\begin{eqnarray}\label{elements-AB}
&&\left( \begin{array}{ccc}
Q_{AA} &Q_{AB} \\\ Q_{BC} &Q_{BB}
\end{array} \right)= \\
&&\left( \begin{array}{ccc}
1\, &\,&-X \\\ 1\, &\,&1-X\\
\end{array} \right)
\left( \begin{array}{ccc}
R_{CC} &R_{CD} \\\ R_{DC} &R_{DD} \\
\end{array} \right)
\left( \begin{array}{ccc}
1 &1 \\\ -X &1-X \\
\end{array} \right)\,\,.\nonumber
\end{eqnarray}
One easily verifies that the Onsager symmetry and sign conditions (\ref{Onsager}) also are valid in the $A,B$ basis.

The last statement seems to contradict a conclusion by Baker and Cahn \cite{Baker-Cahn}, which later has been criticized by Caroli, Caroli, and Roulet \cite{Caroli} as a fallacy, resulting from the assumed absence of solute diffusion in the solid phase. Contrary to this claim, the erroneous conclusion in Ref. \cite{Baker-Cahn} is, to our mind, due to an algebraic mistake, which we here clarify, picking up from Ref. \cite{Baker-Cahn} a model, which has only diagonal elements in the inverse relation of Eq. (\ref{constit-AB}),

\begin{equation}\label{inverse}
J_A=L_{AA}F_A\,\,\,,\,\,\,J_B=L_{BB}F_B\,\,.
\end{equation}
From Eqs. (\ref{AB-fluxes}), (\ref{AB-forces}), (\ref{inverse}) one correctly obtains the first constitutive equation in the $C,D$ basis,

\begin{eqnarray}\label{I_C-F_C}
I_C&=&(L_{AA}+L_{BB})E_C\\&&+[(1-X)L_{BB}-XL_{AA}]E_D\,\,.\nonumber
\end{eqnarray}
In order to attain the second equation, which determines $I_D$, the authors of Ref. \cite{Baker-Cahn} just use the relation (\ref{low-V}) together with the result (\ref{I_C-F_C}). This does not exhaust the full content of Eqs. (\ref{inverse}), and as a result, their mapping from the $A,B$ to the $C,D$ basis is singular instead of being a one-to-one mapping. The correct procedure is, to calculate from Eqs. (\ref{AB-fluxes}) the flux

\begin{equation}\label{I_C-I_D}
I_D=(1-X)J_B-XJ_A\,\,,
\end{equation}
which, together with Eqs. (\ref{inverse}), yields

\begin{eqnarray}\label{I_D-F_D}
I_D&=&[(1-X)L_{BB}-XL_{AA}]E_C\\&&+[(1-X)^2L_{BB}+X^2L_{AA}]E_D\,\,.\nonumber
\end{eqnarray}
Comparing this with Eq. (\ref{I_C-F_C}), one immediately observes that the diagonal matrix elements are positive and that the off-diagonal matrix elements are symmetric. It is a simple exercise, to verify that the determinant, formed with these elements, also is positive, in accordance with our general statement below Eq. (\ref{elements-AB}).

\section{Conclusions}

The model, leading to the exact results (\ref{CD-force-flux}), (\ref{constit-CD}), (\ref{elements-CD}) presumably is the simplest one, which can exactly be evaluated. We also have considered the case of a nonzero diffusion constant $D_S$ in the model equation (\ref{diff-coeff}), which only led to slightly more complicated expressions for the Onsager coefficients.

One of the main insights, drawn from our analysis, concerns the redundancies in the transport coefficients, which appear in the low-velocity regime, if one tries to sustain the formal structure of the Onsager approach. Another conclusion is that, in the discussion of steady states, one should not overlook the possibility of a finite growth velocity, which is clearly compatible with a small driving force. This case is necessarily connected with the appearance of a negative friction coefficient in the crystallization flux-force relation, and completely covered by our analysis.

\acknowledgments

We gratefully acknowledge helpful discussions with Efim Brener. A. L. K. expresses his gratitude to the University of D\"usseldorf for its warm hospitality. This work has been supported by the DFG under Grant No. BA 944/3-3, and by the RFBR under Grant No. N10-02-91332.

\end{document}